\documentclass[11pt,a4paper]{article}
\usepackage{mathtools,cancel}

\usepackage{soul,xcolor}
\usepackage{xcolor}
\usepackage{multirow}
\usepackage{pstricks}
\usepackage{dcolumn}
\usepackage{bm}
\usepackage{latexsym}
\usepackage{dcolumn}
\usepackage[utf8x]{inputenc} 
\usepackage{amsmath}
\usepackage{amsfonts,amssymb}
\usepackage{graphicx,epsfig}
\usepackage{color}
\usepackage{psfrag}
\usepackage{amsthm}
\usepackage{ulem} % compatiple with \sout
\usepackage{soul} % compatiple with \st
\usepackage{flushend}
\usepackage{titlesec}
\usepackage{slashed}
\usepackage{authblk}
\usepackage[font=footnotesize]{caption}
\usepackage[hmarginratio=1:1,top=32mm,columnsep=60pt]{geometry}
\newcommand{\eps}{\varepsilon}
\newcommand{\bea}{\begin{eqnarray}}
\newcommand{\eea}{\end{eqnarray}}
\newcommand{\be}{\begin{equation}}
\newcommand{\ee}{\end{equation}}
\newcommand{\ba}{\begin{array}}
\newcommand{\ea}{\end{array}}

\def\beq{\begin{equation}}
\def\eeq{\end{equation}}
\def\bea{\begin{eqnarray}}
\def\eea{\end{eqnarray}}
\def \cB{{\cal B}}
\def \al{\alpha}
\def\lsim{\leq}
\def\gsim{\geq}
\def\eps{\varepsilon}

\begin{document}
%%%%%%%%%%%%%%%%%%%%%%%%%%%%%%%%%%%%%%%%%%%%%%%%%%%%%%%%%%% Title, Authors and Citations %%%%%%%%%%%%%%%%%%%%%%%%%%%%%%%%%%%%%%%%%%%%%%%%%%%%%%%%%%%%%%%%%%%

\unitlength = 1mm
\begin{flushleft}
%SHIP-HEP-2020-03
%[10mm]
PHENO-2024
\end{flushleft}

\title{\textbf{Constraints on Dark Photon and Dark $Z$ Model Parameters in the $B$ and $K$ Meson Decays}}

\author{\normalsize Ahmed Rashed}
\affil{\small
Department  of Physics,  Shippensburg University of Pennsylvania,\\
 Franklin Science Center, 1871 Old Main Drive, Pennsylvania, 17257, USA
 }
\date{}

%%%%%%%%%%%%%%%%%%%%%%%%%%%%%%%%%%%%%%%%%%%%%%%%%%%%%%%%%%%%%\begin{document}
{\let\newpage\relax\maketitle}

\begin{abstract}
\noindent \normalsize 
The study investigates flavor-changing neutral current 
(FCNC) decays of  $B$ and $K$
mesons in the context of a dark 
$U(1)_D$ model with a dark photon/dark $Z$ mass between 10 MeV and 2 GeV.
While the model improves the fit to certain decay distributions, such as $B \to K^{(*)} \ell^+ \ell^- $ and $ B_s \to \phi \mu^+ \mu^- $, it is ruled out by stringent experimental constraints, including atomic parity violation, $K^+ \to \mu^+ + \text{invisible} $, and $ B_s - \overline{B}_s $ mixing. 
To address these constraints, the model is extended with three modifications; allowing additional invisible decays of $ Z_D $, introducing a direct vector coupling of $ Z_D $ to muons, and including a direct coupling of $ Z_D $ to both muons and electrons, with fine-tuning to cancel the mixing-induced coupling to electrons.
Among these extensions, only the third scenario, involving fine-tuned electron coupling, remains consistent with all experimental constraints.
\end{abstract}
%\tableofcontents

%%%%%%%%%%%%%%%%%%%%%%%%%%%%%%%%%%%%%%%%%%%%%%%%%%%%%%%%%%%%%

\newpage

\section{Introduction}
\label{sec:introduction}
%%%%%%%%%%%%%%%%%%%%%%%%%%%%%%%%%%%
Flavor-changing neutral currents (FCNCs) serve as sensitive probes for new physics (NP) and strongly constrain extensions of the Standard Model (SM). This study examines how recent FCNC measurements in the \( B \)-meson system constrain models involving a light gauge boson, focusing on the dark \( U(1)_D \) model, which predicts either a dark photon or a dark \( Z \) (often referred to as \( Z' \)) depending on its couplings. A dark photon arises solely through kinetic mixing with the electromagnetic field strength~\cite{Holdom:1985ag}, while a dark \( Z \) also involves mass mixing with SM gauge bosons~\cite{Gopalakrishna:2008dv, Davoudiasl:2012ag}. Both scenarios involve loop-induced FCNC processes, with significant contributions from up-type quarks in \( B \) decays due to the large top quark mass, while \( D \) decays are suppressed by down-type quark contributions.

A key feature of models with light mediators is the \( q^2 \)-dependent Wilson coefficients (WCs) they generate, distinguishing them from heavy NP that can be integrated out. The role of light mediators in \( b \to s \ell^+ \ell^- \) transitions has been extensively explored~\cite{Datta:2017pfz, Sala:2017ihs, Bishara:2017pje, Ghosh:2017ber, Datta:2017ezo, Altmannshofer:2017bsz, Datta:2018xty, Datta:2019zca, Darme:2020hpo, Borah:2020swo, Darme:2021qzw, Crivellin:2022obd}. This study focuses on a vector mediator \( Z_D \) with mass \( 0.01 < M_{Z_D} \, \text{GeV} < 2 \), considering both on-shell and off-shell decays. FCNC processes for both dark photon and dark \( Z \) scenarios are analyzed, incorporating leptonic, hadronic, and invisible decays of \( Z_D \). This work also accounts for hadronic decay contributions, often overlooked, and corrects earlier results~\cite{Xu:2015wja} by including the \( q^2 \)-independent dark \( Z \) monopole operator.

The model is extended to include direct couplings of \( Z_D \) to muons, and to both muons and electrons, in addition to mixing-induced couplings. Invisible \( Z_D \) decays to dark sector particles are also considered. Constraints from \( b \to s \ell^+ \ell^- \) data and other low-energy experiments are used to limit the model's parameter space.

\section{Formalism}

The $Z_D$ boson is associated with a broken \( U(1)_D \) gauge symmetry in the dark sector, coupling to the Standard Model (SM) through kinetic mixing with \( U(1)_Y \). The gauge Lagrangian is written as \cite{Holdom:1985ag, Davoudiasl:2012ag}:

\[
\mathcal{L}_\text{gauge} = -\frac{1}{4} B_{\mu\nu} B^{\mu\nu} + \frac{1}{2} \frac{\varepsilon}{\cos\theta_W} B_{\mu\nu} Z_D^{\mu\nu} - \frac{1}{4} Z_{D\mu\nu} Z_D^{\mu\nu}\,,
\]

with \(\varepsilon\) as the kinetic mixing parameter. After diagonalizing the gauge sector, \(Z_D\) couples to the SM via:

\[
\mathcal{L}_D^{\text{em}} \supset e \varepsilon Z_D^\mu J_\mu^{\text{em}} - i e \varepsilon \left[\left[Z_D W^+ W^- \right]\right]\,, 
\]

where the interaction includes both the electromagnetic current and terms involving \(W\)-bosons. If \(U(1)_D\) is broken by a scalar charged under the SM, mass mixing occurs, leading to physical eigenstates expressed as \cite{Gopalakrishna:2008dv, Davoudiasl:2012ag}:

\[
Z = Z^0 \cos\xi - Z_D^0 \sin\xi\,, \quad Z_D = Z^0 \sin\xi + Z_D^0 \cos\xi\,,
\]

where \(\xi\) parameterizes the mixing. The mass mixing induces couplings between \(Z_D\) and SM fields, as described by:

\[
\mathcal{L}_D^Z \supset \frac{g}{\cos \theta_W} \varepsilon_Z Z_D^\mu J_\mu^{Z} - i g \cos \theta_W \varepsilon_Z \left[\left[Z_D W^+ W^- \right]\right]\,,
\]

with \(\varepsilon_Z \equiv \frac{1}{2} \tan 2\xi\). The model parameters are \(\varepsilon\), \(\varepsilon_Z\), and \(M_{Z_D}\), constrained by experimental data \cite{Bertuzzo:2018ftf,Bertuzzo:2018itn,Link:2019pbm}.

Flavor-changing neutral current (FCNC) decays such as \(b \to s Z_D\) occur via loop diagrams involving \(d_i \to d_j Z_D\). The effective Hamiltonian includes monopole and dipole operators with Wilson coefficients:

\[
H_{\rm eff} \supset 
\left(\bar{F}_a^\prime \gamma^\mu P_{L,R} F_b \right) \left[\left(E^{0,c}_{a,b}\right)_{L,R} g^{\mu\nu} + \left(g^{\mu\nu}q^2 - q^\mu q^\nu\right)\left(E^{2,c}_{a,b}\right)_{L,R}\right]V_\nu^c\,,
\]

and

\[
H_{\rm eff} \supset 
\left(\bar{F}_a^\prime \sigma^{\mu\nu} P_{L,R} F_b q_\mu V^c_\nu\right)\left(M^{1,c}_{a,b}\right)_{L,R}\,.
\]

The hadronic part of the amplitude is expressed as:

\begin{eqnarray}
\mathcal{M}_{Z_D} &=&  \langle \mathcal{H}_2 | \bar d_i \gamma_\mu P_{L/R} d_j | \mathcal{\bar H}_1   \rangle  \left[\lbrace\left(E^{0,A}_{c_1,c_2}\right)_{L/R}+\left(E^{0,Z}_{c_1,c_2}\right)_{L/R}\rbrace g^{\mu\nu}\right. \nonumber\\
&+& \left.\lbrace\left(E^{2,A}_{c_1,c_2}\right)_{L/R}+\left(E^{2,Z}_{c_1,c_2}\right)_{L/R}\rbrace \left(g^{\mu\nu}q^2-q^\mu q^\nu\right)\right]V_\nu^{Z_D} \nonumber\\
&+&   \langle \mathcal{H}_2 | \bar d_i iq_\mu \sigma^{\mu\nu} P_{L/R} d_j | \mathcal{\bar H}_1 \rangle  \lbrace\left(M^{1,A}_{c_1,c_2}\right)_{L/R}+\left(M^{1,Z}_{c_1,c_2}\right)_{L/R}\rbrace V_\nu^{Z_D}\,,
\label{DarkPhotonZ}
\end{eqnarray}

Amplitudes for FCNC processes (\(b \to s Z_D, s \to d Z_D, b \to d Z_D\)) are computed using the {\tt Peng4BSM@LO} package \cite{Bednyakov:2013tca}.

Semileptonic \(b \to s \ell^+ \ell^-\) decays involve effective Hamiltonians of the form:

\[
\mathcal{H}_{\text{eff}}^{bs\ell\ell} = -\frac{4 G_F}{\sqrt{2}} \frac{e^2}{16 \pi^2} V_{tb} V_{ts}^* \sum_i (\mathcal{C}_i \mathcal{O}_i + \mathcal{C}_i^\prime \mathcal{O}_i^\prime)\,,
\]

with Wilson coefficients \(\mathcal{C}_{9,\ell}\) and \(\mathcal{C}_{10,\ell}\) given by:

\bea
\mathcal{C}_{9,\ell} &=&\left[\left(\left(E^{0,Z}_{c_1,c_2}\right)_{L} + \lbrace\left(E^{2,A}_{c_1,c_2}\right)_{L}+\left(E^{2,Z}_{c_1,c_2}\right)_{L}\rbrace q^2 \right) \right. \nonumber\\
& \times &  \left. \left(\frac{1}{q^2 - M_{Z_D}^2 + i \Gamma_{Z_D} M_{Z_D}}\right) \left(e\eps + \frac{g}{c_W}\eps_Z g_V^\ell \right) \right]\,, \label{c9}\\
\mathcal{C}_{10,\ell} &=& \left[\left( \left(E^{0,Z}_{c_1,c_2}\right)_{L} + \lbrace\left(E^{2,A}_{c_1,c_2}\right)_{L}+\left(E^{2,Z}_{c_1,c_2}\right)_{L}\rbrace q^2 \right) \right. \nonumber\\
& \times &  \left. \left(\frac{1}{q^2 - M_{Z_D}^2 + i \Gamma_{Z_D} M_{Z_D}}\right) \left(\frac{g}{c_W}\eps_Z g_A^\ell \right) \right]\,, \label{eq:WC-NP}\\
\mathcal{C}_{9,\ell}^{\prime} &=& \mathcal{C}_{10,\ell}^{\prime} = 0\,,
\eea
where $g_V^\ell = (-1 + 4 s^2_W)/2$ and $g_A^\ell = -1/2$ are the vector and axial vector coupling constants for the SM $Z\ell \ell$ interaction; $s_W$ and $c_W$ are the sine and cosine of $\theta_W$, respectively.  When obvious, we suppress the 
$\ell$ subscript in the WCs.

If \(M_{Z_D}\) lies within the kinematic range of the decay, on-shell \(Z_D\) contributions can significantly affect branching fractions \cite{Altmannshofer:2017bsz}.

For \(Z_D\), decays to \(e^+ e^-\), \(\mu^+ \mu^-\), and hadronic final states are considered. Partial widths for leptonic and hadronic decays are computed using vector meson dominance (VMD) models \cite{Tulin:2014tya, Ilten:2018crw, Foguel:2022ppx}, with the hadronic width estimated as:

\[
\Gamma(Z_D \to \mathcal{H}) = \Gamma(Z_D \to \mu^+ \mu^-) \times \mathcal{R}^\mathcal{H}_\mu\,,
\]

where \(\mathcal{R}^\mathcal{H}_\mu\) is derived from experimental data \cite{ParticleDataGroup:2022pth}.

\section{Models}
\label{sec:meth}
We study three different cases of the light $Z_D$ model as specified below.
\begin{itemize}
    \item {\textbf{Case A}}: This is the dark photon and dark $Z$ model. The model has two mixing parameters ($\eps$ and $\eps_Z$) and the mass $M_{Z_D}$.
    \item {\textbf{Case B}}: A muonphilic $Z_D$ in which Case A is extended with an additional direct interaction of the dark $Z$ with muons:
    \beq 
\mathcal{L}_D^Z \supset g_{D}^\mu \bar{\mu} \gamma_\al \mu Z_{D}^\al\,.
\label{eq:CaseB}
\eeq 
This scenario has an additional free parameter $g_{D}^\mu$.
    \item {\textbf{Case C}}: Case A is extended with additional direct interactions of the dark $Z$ with both electrons and muons: 
    \beq 
\mathcal{L}_D^Z \supset g_{D}^e \bar{e} \gamma_\al e Z_{D}^\al + g_{D}^\mu \bar{\mu} \gamma_\al \mu Z_{D}^\al\,.
\label{eq:CaseC}
\eeq
We assume $g_{D}^e$ is fine-tuned so that it cancels the coupling of $Z_D$ to electrons via mixing. Then, all observables for the electron mode are described by the SM only.
\end{itemize}

%%%%%%%%%%%%%%%%%%%%%%%%%%%%%%%%% 
\section{Constraints}

%%%%%%%%%%%%%%%%%%%%%%%%%%%%
\subsection{$B_s$ mixing}
%%%%%%%%%%%%%%%%%%%%%%%%%%%%%%%%%%%%%

$B$ meson mixing is a significant tool for probing new physics, providing stringent constraints on theoretical models. In the Standard Model (SM), the mixing originates from a box diagram involving a $W$ boson and a top quark~\cite{Artuso:2015swg}. The dominant SM contribution to the mass difference between $B_s^0$ and $\bar{B}_s^0$ mesons depends on QCD corrections, the top quark mass~\cite{FermilabLattice:2016ipl}, and hadronic parameters like the decay constant and bag parameter~\cite{Buchalla:1995vs}. Lattice QCD calculations have refined these parameters~\cite{Aoki:2021kgd}.

New Physics (NP) contributions, such as from a potential dark $Z_D$ boson, modify the mass difference. These contributions depend on the $Z_D$ mass and coupling constants~\cite{Datta:2017ezo}. The experimental mass difference aligns closely with SM predictions~\cite{DiLuzio:2019jyq, ParticleDataGroup:2022pth}, placing tight constraints on NP parameters. Notably, lighter $Z_D$ bosons require smaller couplings to remain consistent with observations, disallowing certain parameter combinations, particularly for $Z_D$ masses below 60 MeV and couplings above 0.001. These findings underscore the sensitivity of $B_s^0 - \bar{B}_s^0$ mixing to new physics.

\begin{figure}[h]
    \centering
    \includegraphics[scale=0.85]{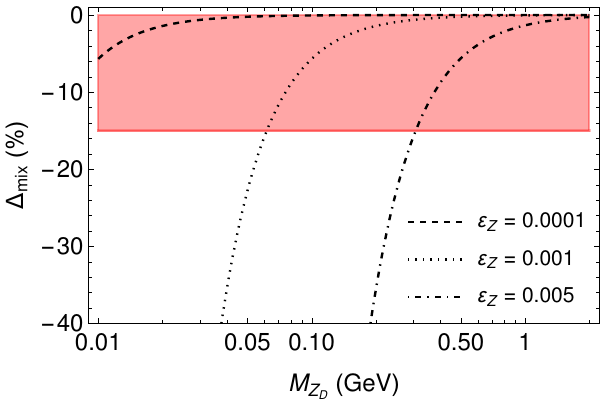}
    \caption{Sensitivity of $B_s^0-\overline{B}_s^0$ mixing to $\eps_Z$ as a function of $M_{Z_D}$. At leading order $\Delta_{mix}$ is independent of $\eps$. The red band is the uncertainty in $\Delta_{\rm mix}$ taken to be the $2\sigma$ lower uncertainty in $\Delta M_{B_s}^{SM}$.}
    \label{fig:Bs-mixing-CaseA}
\end{figure}

We plot $\Delta_{mix}$ as a function of the dark $Z$ mass for different values of $\eps_Z$ in Fig.~\ref{fig:Bs-mixing-CaseA}. It is evident that lighter $Z_D$ require smaller values of $\eps_Z$ for $\Delta_{mix}$ to lie within the $2\sigma$ uncertainty of the SM prediction. 
We find that $\eps_Z 
\gsim 0.001$ is disallowed for 
$M_{Z_D} \lsim 60$ MeV.

%%%%%%%%%%%%%%%%%%%%%%%%%%%%%%%%%%%%%%%%%%%%%
\subsection{$B_s \to \mu^+ \mu^-$}
%%%%%%%%%%%%%%%%%%%%%%%%%%%%%%%%%%%%%%%%%%%%%
The rare decay $B_s \to \mu^+\mu^-$ is a crucial probe for new physics. Its branching fraction depends on Standard Model (SM) parameters and possible contributions from new physics. In the $Z_D$ model, contributions from scalar and pseudoscalar Wilson coefficients are absent, leaving the dominant effect from $\mathcal{C}_{10}$~\cite{FlavourLatticeAveragingGroup:2019iem}. 

The SM prediction for the branching fraction is $\mathcal{B}(B_s^0 \to \mu^+\mu^-)^{SM} = (3.67 \pm 0.15) \times 10^{-9}$~\cite{Altmannshofer:2021qrr, Guadagnoli:2022izc}, while the experimental measurement from LHCb is $\mathcal{B}(B_s^0 \to \mu^+\mu^-)^{LHCb} = (3.09^{+0.46+0.15}_{-0.43-0.11}) \times 10^{-9}$~\cite{LHCb:2021awg}. The close agreement between these values constrains potential new physics contributions, highlighting the importance of precision in both theoretical predictions and experimental measurements.

\begin{figure}[h]
    \centering
    \includegraphics[scale=0.85]{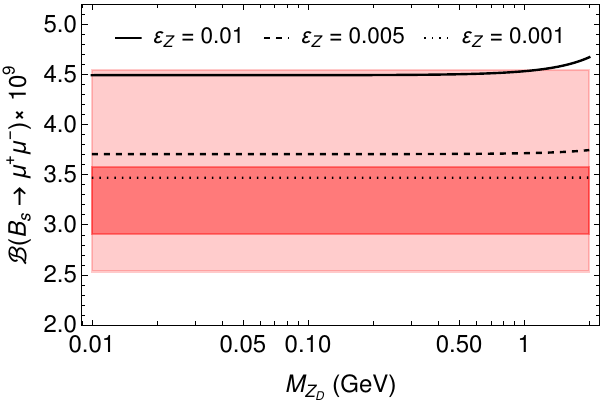}
    \caption{Branching fraction for $B_s \to \mu^+ \mu^-$ for different values of $\varepsilon_Z$. The horizontal red (light red) band denotes the $1\sigma\, (3 \sigma)$ allowed region from experiment~\cite{LHCb:2021awg}. The decay rate does not depend on $\eps$.} 
    \label{fig:BsMuMu}
\end{figure}

The $Z_D$ contribution to this rare decay is shown in Fig.~\ref{fig:BsMuMu}. Since the decay rate depends only on the new axial-vector interaction of the dark $Z$, it is independent of 
$\eps$. It is evident that $\eps_Z$ as large as $0.01$ is allowed by the data at the $3\sigma$ confidence level (CL). 

%%%%%%%%%%%%%%%%%%%%%%%%%%%%%%%%%%%%%%%%%%%%%%%
\subsection{$B \to K^{(*)} \nu \bar{\nu}$}
%%%%%%%%%%%%%%%%%%%%%%%%%%%%%%%%%%%%%%%%%%%%%%%

The decays $B \to K^{(*)} \nu \bar{\nu}$ are sensitive probes for new physics, particularly in the context of the $Z_D$ model. These decays involve the on-shell production of $Z_D$ in $B$ decay, followed by $Z_D \to \nu \bar{\nu}$~\cite{Fuyuto:2015gmk}. The effective interaction strengths and form factors governing these decays are derived from loop functions, and their detailed expressions are available in~\cite{Ball:2004rg}.

The Standard Model (SM) predicts branching fractions of $\mathcal{B}(B^+ \to K^+ \nu \bar{\nu})_{SM} = (4.4 \pm 0.7) \times 10^{-6}$ and $\mathcal{B}(B^0 \to K^{*0} \nu \bar{\nu})_{SM} = (11.6 \pm 1.1) \times 10^{-6}$~\cite{Felkl:2021uxi}. A recent search by Belle II for $B^+ \to K^+ \nu \bar{\nu}$ sets a $90\%$ confidence level (CL) upper bound of $\mathcal{B}(B^+ \to K^+ \nu \bar{\nu}) < 4.1 \times 10^{-5}$~\cite{Dattola:2021cmw}, and the weighted average from existing data is $\mathcal{B}(B^+ \to K^+ \nu \bar{\nu})_{WA} = (1.1 \pm 0.4) \times 10^{-5}$~\cite{Felkl:2021uxi}, showing a slight enhancement over the SM prediction. However, this enhancement should not yet be interpreted as evidence for new physics.

For $B \to K^* \nu \nu$, the most recent $90\%$ CL upper bound from Belle is $\mathcal{B}(B^0 \to K^{*0} \nu \bar{\nu}) < 1.8 \times 10^{-5}$~\cite{Belle:2017oht}. These results impose constraints on new physics scenarios while leaving room for further exploration.

\begin{figure}[h]
    \centering
    \includegraphics[scale=0.65]{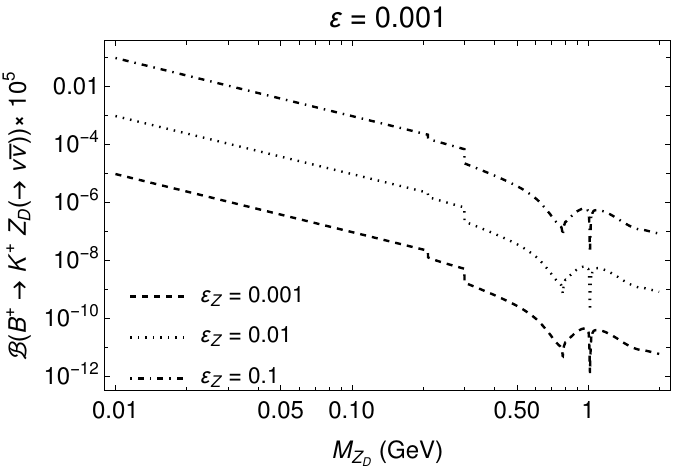}~~
    \includegraphics[scale=0.65]{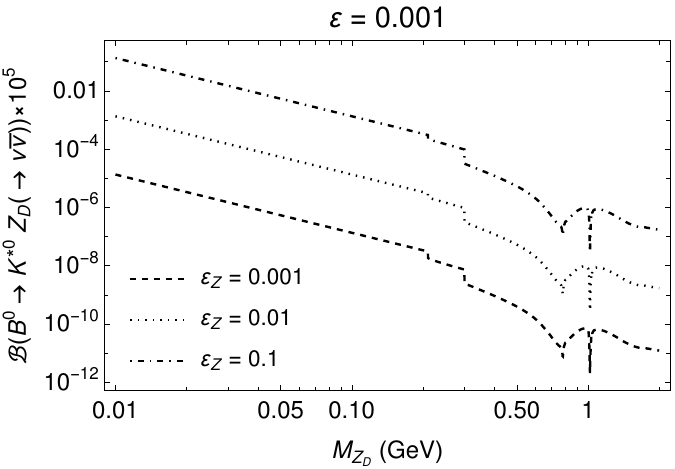}
    \caption{Branching fraction of $B \to K^{(*)} \nu \bar\nu$ as a function of the $Z_D$ mass for three values of $\eps_Z$ and $\eps=0.001$.}
    \label{fig:BtoKnunu}
\end{figure}

In Fig.~\ref{fig:BtoKnunu}, we plot the branching fractions for some benchmark values of $\eps_Z$. We find that they are more than an order of magnitude smaller than the respective upper bounds even for $\eps_Z$ as large as 0.1. 

%%%%%%%%%%%%%%%%%%%%%%%%%%%%%%%%%%%%%%%%%%%%%%%
\subsection{Kaon decay and mixing}
%%%%%%%%%%%%%%%%%%%%%%%%%%%%%%%%%%%%%%%%%%%%%%%

The flavor-changing decays $K \to \pi \nu \bar{\nu}$ are governed by the $s \to d \nu \bar{\nu}$ transition. The key decay modes are $K^+ \to \pi^+ \nu \bar{\nu}$ and $K_L \to \pi^0 \nu \bar{\nu}$. The most recent measurement from the NA62 experiment gives $\mathcal{B}(K^+ \to \pi^+ \nu \bar{\nu}) = (10.6^{+4.0}_{-3.4} \pm 0.9) \times 10^{-11}$~\cite{NA62:2021zjw}, while the KOTO experiment places a 90\% confidence level (CL) upper bound of $\mathcal{B}(K_L \to \pi^0 \nu \bar{\nu}) < 4.9 \times 10^{-9}$~\cite{KOTO:2020prk}. These results align with the Standard Model (SM) predictions of $\mathcal{B}(K^+ \to \pi^+ \nu \bar{\nu})_{SM} = (8.4 \pm 1.0) \times 10^{-11}$ and $\mathcal{B}(K_L \to \pi^0 \nu \bar{\nu})_{SM} = (3.4 \pm 0.6) \times 10^{-11}$~\cite{Buras:2015qea,PDG2022}. The two modes are related through the Grossman-Nir bound, which limits $\mathcal{B}(K_L \to \pi^0 \nu \bar{\nu})$ based on $\mathcal{B}(K^+ \to \pi^+ \nu \bar{\nu})$~\cite{Grossman:1997sk}.

In the context of new physics, including the $Z_D$ model, contributions to these decays are resonant but highly suppressed due to the weak $s \to d Z_D$ transition. Consequently, the branching fractions remain small even for relatively large $Z_D$ coupling parameters ($\epsilon_Z \sim 0.01$).

Neutral kaon oscillations ($K^0 \leftrightarrow \overline{K}^0$) are also relevant, with a mass difference $\Delta M_K$ that matches the SM prediction of $3.484(6) \times 10^{-12}~\mathrm{MeV}$~\cite{FlavourLatticeAveragingGroup:2019iem}. The dominant SM contribution arises from loop-level processes with QCD corrections and hadronic parameters, while new physics contributions from the $Z_D$ model, for example, are found to be negligible under typical parameter values, such as $M_{Z_D} = 10~\mathrm{MeV}$ and $\epsilon_Z = 0.001$. Thus, $\Delta M_K$ does not strongly constrain this model.

\subsection{Radiative $K^+ \to \mu^+ \nu_\mu Z_D$ decays}
The three-body decay \( K^+ \to \mu^+ \nu_\mu Z_D \) serves as a radiative correction to the standard \( K^+ \to \mu^+ \nu_\mu \) decay, where the dark boson \( Z_D \) is emitted from the muon leg and subsequently decays invisibly. This decay is relevant for scenarios where the dark boson mass is below \( 2m_\mu \), allowing it to be produced on-shell. Experimental constraints on this process have been established by the NA62 experiment, which sets a 90\% confidence level upper limit on the branching fraction for \( K^+ \to \mu^+ \nu \nu \bar{\nu} \) at \( 1.0 \times 10^{-6} \). This bound imposes restrictions on the coupling parameters of the dark boson in various model scenarios.

\begin{figure}[h]
    \centering
    \includegraphics[scale=0.655]{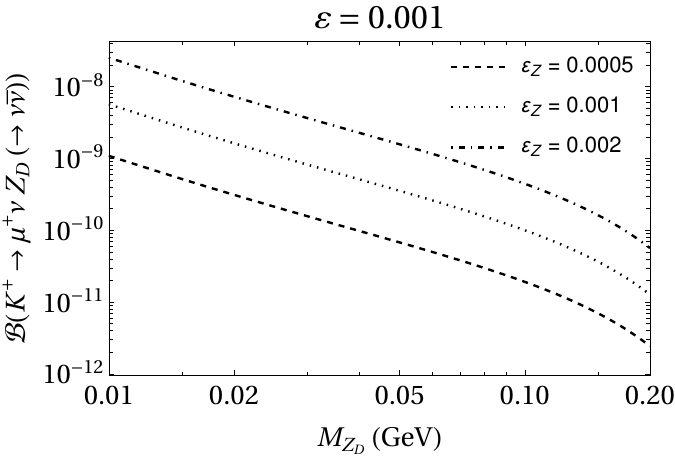}~~
    \includegraphics[scale=0.645]{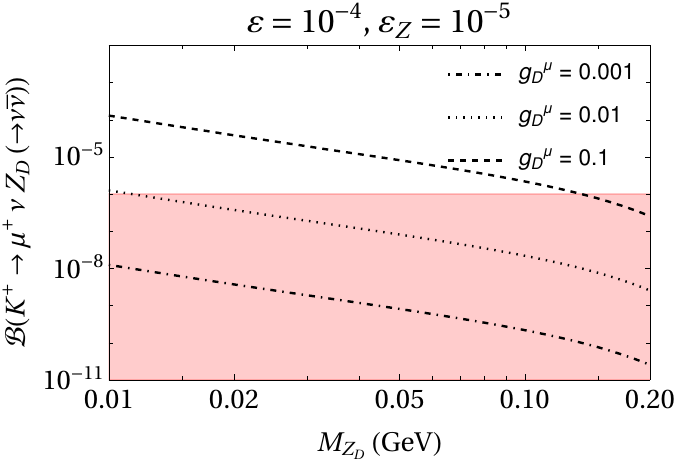}
    \caption{Dependence of the $K^+ \to \mu^+ + invisible$ branching fraction on the mixing parameters in Case A (left) and on the direct coupling $g_{D}^\mu$ in Case B (right). The red shaded region shows the 90\% CL upper limit on the branching fraction.}
    \label{fig:KmunuX}
\end{figure}

In one scenario, referred to as Case A, where the emission is suppressed by mixing parameters, the branching fraction remains below the experimental limit even for relatively large couplings, as illustrated in the left panel of Fig.~\ref{fig:KmunuX}. Conversely, in Case B, where the direct coupling \( g_D^\mu \) dominates, the branching fraction increases with \( g_D^\mu \), necessitating \( g_D^\mu < 0.01 \) to remain consistent with the data. This behavior is shown in the right panel of Fig.~\ref{fig:KmunuX}. The interplay between mixing and direct coupling highlights distinct constraints depending on the underlying assumptions of the model.

\begin{figure}[h]
	\centering
	\includegraphics[scale=0.65]{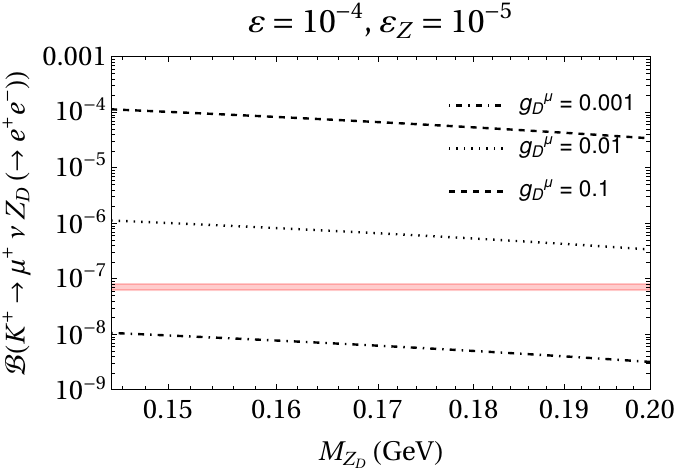}
	\caption{Dependence of the $K^+ \to \mu^+ \nu e^+ e^-$ branching fraction on $g_{D}^\mu$ in Case B. The red shaded region shows the $3\sigma$ interval of the branching fraction.}
    \label{fig:KmunuXee}
\end{figure}

In addition to invisible decays, the dark boson \( Z_D \) can also decay into visible final states, such as \( e^+e^- \). This contributes to the decay \( K^+ \to \mu^+ \nu e^+ e^- \), for which the experimentally measured branching fraction is \( (7.06 \pm 0.31) \times 10^{-8} \) for invariant electron-positron masses above \( 145 \,\mathrm{MeV} \). Within the context of Case B, the coupling \( g_D^\mu \) can enhance the branching fraction significantly for \( 145 < M_{Z_D} < 200 \,\mathrm{MeV} \) when \( \epsilon_Z < \epsilon \) and \( Z_D \) predominantly decays to \( e^+e^- \). This dependence is illustrated in Fig.~\ref{fig:KmunuXee}, where the constraints from data are clearly visible. Notably, these constraints do not apply to scenarios like Case C, where \( Z_D \) lacks coupling to electrons.

Overall, the analysis of both invisible and visible decay modes of the dark boson provides stringent experimental constraints on its properties, demonstrating how rare kaon decays can serve as sensitive probes of new physics.

\subsection{Radiative $\pi^+ \to \mu^+ \nu_\mu Z_D$ decays}
\label{sec:PiMuNu}
\begin{figure}[h]
\centering
\includegraphics[scale=0.65]{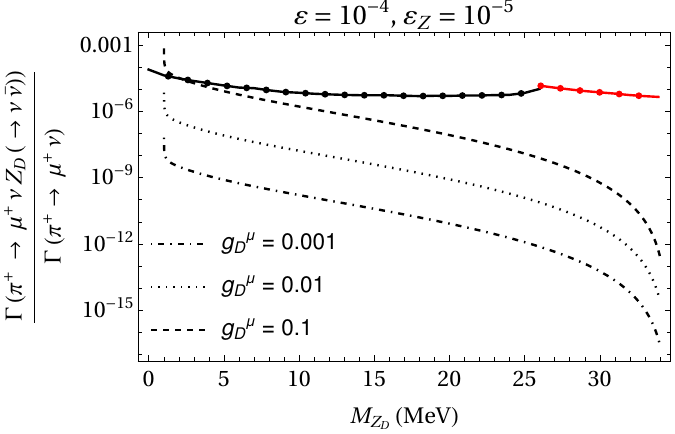}~~
\includegraphics[scale=0.65]{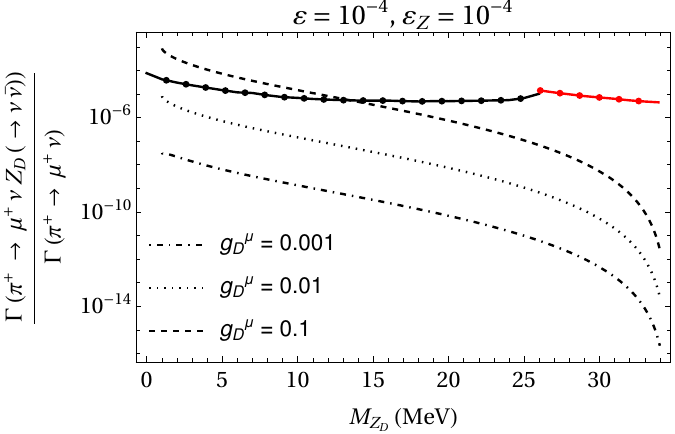}
    \caption{Dependence of $R^{\pi \mu\nu X} = \Gamma(\pi^+ \to \mu^+ \nu_\mu Z_D (\to \nu \bar{\nu}))/\Gamma(\pi^+ \to \mu^+ \nu_\mu)$ on $g_D^\mu$ for $\eps_Z = 10^{-5}$ (left) and $10^{-4}$ (right). The black and red solid curves punctuated with points show the 90\%~CL upper limit from PIENU for the muon kinetic energy ranges, $T_\mu > 1.2$~MeV and $T_\mu < 1.2$~MeV, respectively.  }
\label{fig:PiMuNuX}
\end{figure}

The decay process \( \pi^+ \to \mu^+ \nu_\mu Z_D \), analogous to radiative kaon decay, can be enhanced when the dark boson \( Z_D \) mass satisfies \( 0 < M_{Z_D} < m_\pi - m_\mu \). The PIENU experiment provides constraints on this process by setting an upper bound on the ratio \( R^{\pi \mu\nu X} = \Gamma(\pi^+ \to \mu^+ \nu_\mu X)/\Gamma(\pi^+ \to \mu^+ \nu_\mu) \), where \( X \) is an invisible decay product, in the mass range \( 0 < M_{X} < 33.9 \,\mathrm{MeV} \)perimental bound is used to constrain the parameters of the dark boson model.

The decay rate depends on the coupling structure. In Case A, it is suppressed by mixing parameters, while in Cases B and C, it scales with \( (g_D^\mu)^2 \). The amplitude squared for the process is derived from the general interaction and is adjusted for pion decay by substituting the relevant constants (e.g., \( f_K \to f_\pi \), \( V_{us} \to V_{ud} \), and \( m_K \to m_\pi \)).

Fig.~\ref{fig:PiMuNuX} illustrates the relationship between \( R^{\pi \mu\nu X} \) and \( g_D^\mu \), with experimental limits from PIENU shown for two ranges of muon kinetic energy: \( T_\mu > 1.2 \,\mathrm{MeV} \) (solid black curve) and \( T_\mu < 1.2 \,\mathrm{MeV} \) (solid red curve). The results indicate that \( g_D^\mu < 0.1 \) is consistent with the data for \( M_{Z_D} < 34 \,\mathrm{MeV} \) if the dark \( Z \) does not predominantly decay to neutrinos (i.e., \( \epsilon_Z \ll \epsilon \)). However, for \( \epsilon_Z \gsim \epsilon \), where \( \cB(Z_D \to \nu \bar{\nu}) \) is significant, \( g_D^\mu \geq 0.1 \) is excluded for \( M_{Z_D} < 15 \,\mathrm{MeV} \). This highlights the sensitivity of \( \pi^+ \to \mu^+ \nu_\mu Z_D \) decays to the coupling parameters and branching ratios of the dark boson.

\subsection{Atomic parity violation}
\label{sec:APV}
The dark boson \( Z_D \) can couple to first-generation Standard Model (SM) fermions via mixing, leading to stringent constraints from atomic parity-violating (APV) observables. These interactions affect the weak charge \( Q_W \) of the proton and certain nuclei, such as cesium (\( ^{133}C_s \)), whose measured values are \( Q_W^{p,exp} = 0.0719(45) \) and \( Q_W^{^{133}C_s,exp} = -72.82(42) \), respectivelynfluenced by modifications to the Fermi constant \( G_F \) and the weak mixing angle \( \theta_W \), both of which are altered by the dark boson interaction. The parameters \( \rho_d \) and \( \kappa_d \) encode these modifications, depending on the mixing parameters (\( \epsilon, \epsilon_Z \)), the mass of the dark boson (\( M_{Z_D} \)), and the momentum transfer (\( Q^2 \)). For example, \( \rho_d \) and \( \kappa_d \) involve terms proportional to mixing coefficients and the ratio of \( M_{Z_D} \) to the SM \( Z \)-boson mass. 

The function \( f(Q^2/M_{Z_D}^2) \), which characterizes the momentum dependence, varies for different systems. For protons, \( f \) decreases with increasing \( Q^2 \), while for cesium, it is approximately constant. For instance, \( f \approx 0.5 \) at \( M_{Z_D} \approx 2.4 \,\mathrm{MeV} \) and \( f \approx 1 \) for \( M_{Z_D} \approx 100 \,\mathrm{MeV} \). Using these dependencies, the model's consistency with experimental data from APV measurements provides constraints on the mixing parameters and the mass of \( Z_D \).
\begin{figure}[h]
    \centering
    \includegraphics[scale=0.75]{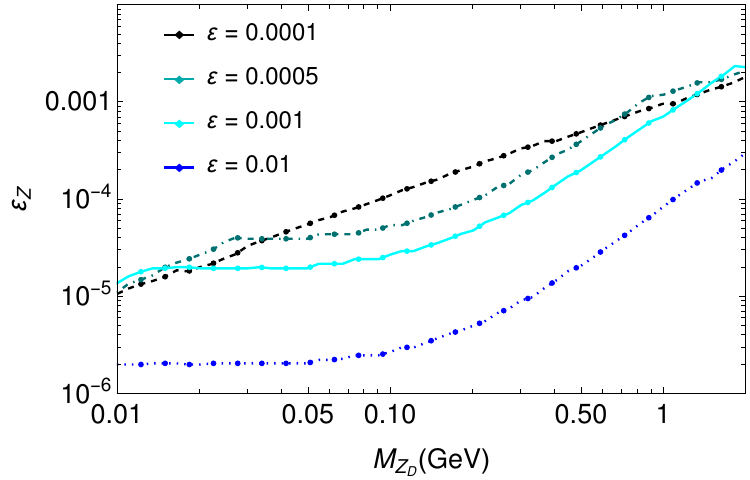}
    \caption{The $3\sigma$ CL upper bound on $\eps_Z$ from measurements of the proton and cesium weak charges in atomic parity violation experiments.}
    \label{fig:APV}
\end{figure}

In Fig.~\ref{fig:APV}, we plot the $3\sigma$ CL upper bound from APV on $\eps_Z$  for different values of $\eps$. By and large, for larger values of $\eps$, 
$\eps_Z$ is more constrained. 
Among the constraints discussed so far, APV places the strongest constraint on $\eps_Z$ in the few MeV-GeV mass range. The coupling $g_{D}^\mu$ which appears in Case B is unconstrained by APV. Again, because of our fine-tuned choice of $g_{D}^e$ to cancel 
the $Z_D$ coupling to electrons,
Case C is also unconstrained by APV.

\subsection{Neutrino trident and CE$\nu$NS}

Muon neutrinos can scatter off a nucleus and produce a pair of muons via a weak interaction known as neutrino trident production, which occurs through \( Z_D \) exchange. This process has been measured in neutrino beam experiments, such as CHARM-II and CCFR, with results indicating a ratio of the experimental to Standard Model cross sections close to 1 for both experiments. The absence of any excess in these measurements places strong upper limits on the coupling of \( Z_D \) with muons. In particular, bounds derived for a \( Z' \) based on the \( U(1)_{L_\mu - L_\tau} \) symmetry are adapted to constrain the model’s parameters in the \( M_{Z_D} \) and \( \epsilon_Z \) plane, with the bounds being less stringent for Case A, where both vector and axial-vector interactions are involved.

Additionally, the COHERENT experiment, which observed coherent elastic neutrino-nucleus scattering (CE$\nu$NS), provides further constraints on the model. The data from COHERENT set limits on the \( \epsilon_Z \) parameter for a given mass of \( Z_D \). These bounds, derived by rescaling earlier results from \( U(1)_X \) models, turn out to be much weaker compared to direct gauge coupling constraints. For instance, for \( M_{Z_D} \sim 10 \, \text{MeV} \), the upper bound on \( \epsilon_Z \) is found to be approximately 0.0005.

\subsection{Collider and other bounds}
\label{sec:collider-bounds}

The dark boson \( Z_D \) can be produced through both on-shell and off-shell decays of the \( Z \) boson, such as in the process \( Z \to \ell \ell Z_D \), which could lead to final states like \( Z \to 4\ell \). Searches for such events at ATLAS and CMS have shown results consistent with the Standard Model, placing a lower bound on \( M_{Z_D} \) of about 5 GeV. However, due to suppression of the \( Z_D \)-lepton coupling in the model, the decay rate for \( Z_D \to 4\ell \) is small, and the resulting bounds are not impactful for this model.

Belle II has also conducted a search for invisibly decaying \( Z^\prime \) bosons in the \( e^+ e^- \to \mu^+ \mu^- \) plus missing energy channel, placing an upper limit on the coupling of \( Z^\prime \) to muons for \( M_{Z'} < 6 \, \text{GeV} \). However, this bound is weaker than those from low-energy experiments and does not significantly constrain the model.

The dark boson could contribute to the leptonic decay width of the \( W \) boson, particularly in processes where the \( Z_D \) decays invisibly. The contribution to the decay width is found to be small enough to be consistent with the measured \( W \) decay width. For Case A, assuming \( M_{Z_D} \) is below the dimuon threshold, the contribution is constrained to be within the uncertainty of the measured \( W \) width, setting an upper limit on \( \epsilon_Z \) for different values of \( M_{Z_D} \). This limit, however, is not as stringent as the other constraints discussed.

Additionally, LHCb searches for dark photons in dimuon samples have set strong bounds on the mixing parameter \( \epsilon \). For \( M_{Z_D} \sim 200 \, \text{MeV} \), the bound on \( \epsilon \) is \( \sim 10^{-4} \), and for \( M_{Z_D} \sim 2 \, \text{GeV} \), it is \( \sim 0.0005 \), which directly applies to Case A. For Cases B and C, this bound is recast in terms of the \( g_D^\mu \) coupling, which rules out \( g_D^\mu > 10^{-3} \) for \( M_{Z_D} > 210 \, \text{MeV} \).

\section{Parameter fits}

\begin{table}[t]
\centering
\footnotesize
\renewcommand{\arraystretch}{1.2}
\resizebox{\columnwidth}{!}{\begin{tabular}{|c|c|c|c|c|}
\hline
\textbf{Decay} & \textbf{Ref.} & $\mathbf{q^2}$ \textbf{bin (GeV$^2$)}  &  \textbf{Measurement} & \textbf{SM expectation} \\
\hline
\multirow{4}{*}{$\frac{d\mathcal{B}}{dq^2}(B^0 \to K^{*0} \mu^+ \mu^-) \times 10^{8}$} & \multirow{4}{*}{\cite{LHCb:2016ykl}} & $0.1-0.98$ & $11.06^{+0.67}_{-0.73}\pm 0.29 \pm 0.69$ & $10.60 \pm 1.54$ \\
& & $1.1-2.5$ & $3.26^{+0.32}_{-0.31}\pm 0.10 \pm 0.22$ & $4.66\pm 0.74$ \\
& & $2.5-4.0$ & $3.34^{+0.31}_{-0.33}\pm 0.09 \pm 0.23$ & $4.49\pm 0.70$ \\
& & $4.0-6.0$ & $3.54^{+0.27}_{-0.26}\pm 0.09 \pm 0.24$ & $5.02\pm 0.75$ \\
\cline{1-5}
\multirow{3}{*}{$\frac{d\mathcal{B}}{dq^2}(B^+ \to K^{*+} \mu^+ \mu^-) \times 10^{8}$} & \multirow{3}{*}{\cite{LHCb:2014cxe}} & $0.1-2.0$ & $5.92^{+1.44}_{-1.30}\pm 0.40$ & $7.97\pm 1.15$ \\
& &  $2.0-4.0$ & $5.59^{+1.59}_{-1.44}\pm 0.38$ & $4.87\pm 0.76$ \\
& & $4.0-6.0$ & $2.49^{+1.10}_{-0.96}\pm 0.17$ & $5.43\pm 0.74$ \\
\cline{1-5}
\multirow{6}{*}{$\frac{d\mathcal{B}}{dq^2}(B^+ \to K^{+} \mu^+ \mu^-) \times 10^{8}$} & \multirow{6}{*}{\cite{LHCb:2014cxe}} & $0.1-0.98$ & $3.32\pm 0.18 \pm 0.17$ & $3.53 \pm 0.64 $ \\
& & $1.1-2.0$ & $2.33\pm 0.15 \pm 0.12$ & $3.53 \pm 0.58$ \\
& & $2.0-3.0$ & $2.82\pm 0.16 \pm 0.14$ & $3.51 \pm 0.52$ \\
& & $3.0-4.0$ & $2.54\pm 0.15 \pm 0.13$ & $3.50 \pm 0.63$ \\
& & $4.0-5.0$ & $2.21\pm 0.14 \pm 0.11$ & $3.47 \pm 0.60$ \\
& & $5.0-6.0$ & $2.31\pm 0.14 \pm 0.12$ & $3.45 \pm 0.53$ \\
\cline{1-5}
\multirow{3}{*}{$\frac{d\mathcal{B}}{dq^2}(B^0 \to K^{0} \mu^+ \mu^-) \times 10^{8}$} & \multirow{3}{*}{\cite{LHCb:2014cxe}} & $0.1-2.0$ & $1.22^{+0.59}_{-0.52} \pm 0.06$ & $3.28 \pm 0.52$ \\
& & $2.0-4.0$ & $1.87^{+0.55}_{-0.49} \pm 0.09$ & $3.25 \pm 0.56$ \\
& & $4.0-6.0$ & $1.73^{+0.53}_{-0.48} \pm 0.09$ & $3.21 \pm 0.54$ \\
\cline{1-5}
\multirow{4}{*}{$\frac{d\mathcal{B}}{dq^2}(B_s^0 \to \phi \mu^+ \mu^-) \times 10^{8}$} & \multirow{4}{*}{\cite{LHCb:2021zwz}} & $0.1-0.98$ & $7.74 \pm 0.53 \pm 0.12 \pm 0.37$ & $11.31 \pm 1.34$ \\
& & $1.1-2.5$ & $3.15\pm 0.29 \pm 0.07 \pm 0.15$ & $5.44 \pm 0.61$ \\
& & $2.5-4.0$ & $2.34 \pm 0.26 \pm 0.05 \pm 0.11$ & $5.14 \pm 0.73$ \\
& & $4.0-6.0$ & $3.11 \pm 0.24\pm 0.06 \pm 0.15$ & $5.50 \pm 0.69$ \\
\cline{1-5}
\multirow{2}{*}{$\mathcal{B}(B^+ \to K^{+} e^+e^-) \times 10^{8}$} & \multirow{2}{*}{\cite{BELLE:2019xld}} & $0.1-4.0$ & $18.0^{+3.3}_{-3.0} \pm 0.5$ & $13.73 \pm 1.88$ \\
& & $4.0-8.12$ & $9.6^{+2.4}_{-2.2} \pm 0.3$ & $14.11 \pm 1.88$ \\
%& & $1.0-6.0$ & $16.6^{+3.2}_{-2.9} \pm 0.4$ & $17.45 \pm 3.03$ \\
\cline{1-5}
$\mathcal{B}(B^0 \to K^{*0}e^+ e^-) \times 10^{7}$ & \cite{LHCb:2013pra} &$0.03^2-1.0^2$ & $3.1^{+0.9+0.2}_{-0.8-0.3} \pm 0.2$ & $2.56 \pm 0.44$ \\
\cline{1-5}
$\mathcal{B}(B \to X_s \mu^+ \mu^-) \times 10^{6}$ & \cite{BaBar:2013qry} & $1.0-6.0$ & $0.66^{+0.82+0.30}_{-0.76-0.24} \pm 0.07$ & $1.67 \pm 0.15$ \\
$\mathcal{B}(B \to X_s e^+ e^-) \times 10^{6}$ & \cite{BaBar:2013qry} & $1.0-6.0$ & $1.93^{+0.47+0.21}_{-0.45-0.16} \pm 0.18$ & $1.74 \pm 0.16$ \\
\hline
$\frac{d\mathcal{B}}{dq^2}(B^+ \to K^{+} e^+e^-) \times 10^{9}$ & \cite{LHCb:2022zom} & $1.1-6.0$ & $25.5^{+1.3}_{-1.2}\pm 1.1$ & $34.9\pm 6.2$\\
\hline
$\frac{d\mathcal{B}}{dq^2}(B^0 \to K^{*0} e^+e^-) \times 10^{9}$ & \cite{LHCb:2022zom} & $1.1-6.0$ & $33.3^{+2.7}_{-2.6}\pm 2.2$ & $47.7\pm 7.5$\\
\hline
\end{tabular}}
\caption{Experimental measurements and SM expectations in $q^2$ bins. The SM $\chi^2$ for the fit to all the observables is 93.56, and for just the muon modes it is 84.30.}
\label{tab:data}
\end{table}

The study fits recent experimental data on exclusive decays (\( B \to K^{(*)} \ell^+ \ell^- \), \( B_s^0 \to \phi \mu^+ \mu^- \)) and inclusive decays (\( B \to X_s \ell^+ \ell^- \)) in different \( q^2 \) bins. The fits use the software \texttt{flavio} to compute both Standard Model (SM) and New Physics (NP) predictions, with the best fit values determined by minimizing the chi-squared function over the experimental data. The results are displayed in terms of $1\sigma$, $2\sigma$, and $3\sigma$ confidence level (CL) regions for the model parameters. The analysis excludes data below \( c \bar{c} \) resonances.

In Case A, the model is constrained by bounds on the mixing parameters, \( \epsilon \) and \( \epsilon_Z \), with the best fit occurring at \( M_{Z_D} = 10.07 \, \text{MeV} \), \( \epsilon = 1.6 \times 10^{-5} \), and \( \epsilon_Z = 0.002 \). The allowed regions for the parameters are shown in two-dimensional plots for \( (M_{Z_D}, \epsilon_Z) \) and \( (M_{Z_D}, \epsilon) \), with the best fit marked by a blue circle. However, the model faces tensions with low-energy constraints, particularly from atomic parity violation (APV) experiments, which exclude parameter space for \( M_{Z_D} \lesssim 30 \, \text{MeV} \) at more than $3\sigma$. Further consideration of the dark boson’s invisible decay width shows negligible impact on the fit, and the best fit remains unchanged.

In Case B, the direct interaction of the dark \( Z \) with muons improves the fit significantly, with the best fit occurring at \( M_{Z_D} = 10.3 \, \text{MeV} \) and \( g_D^\mu = 0.28 \). Despite these improvements, the model is still excluded by measurements such as \( K^+ \to \mu^+ \nu X \) and the \( W \)-boson width. The allowed parameter space is ruled out by these experimental constraints, as shown in the figure for Case B. A further extension with an axial-vector coupling of the dark \( Z \) to muons (Case C) leads to additional contributions, but the constraints from kaon decays and leptonic \( W \) decays impose even stricter limits, making this possibility less favorable.

Case C considers a direct coupling of the dark boson to electrons, allowing the model to bypass APV constraints by fine-tuning the coupling to cancel the \( Z_D \)-electron interaction via mixing. The best fit point occurs at \( M_{Z_D} = 30.2 \, \text{MeV} \) and \( g_D^\mu = 0.033 \). This scenario provides a marked improvement over Case B, with a better fit to the binned \( b \to s \mu^+ \mu^- \) data and an order of magnitude smaller \( g_D^\mu \). The parameter space remains consistent with bounds from neutrino trident production, \( K \to \mu + \text{invisible} \), and the \( W \)-boson width. The allowed region, which fits the data at $2\sigma$ CL, is also shown in the figure for Case C, where dark photon searches at LHCb further constrain the parameter space.

\begin{figure}[h]
    \centering
    \includegraphics[scale=0.4]{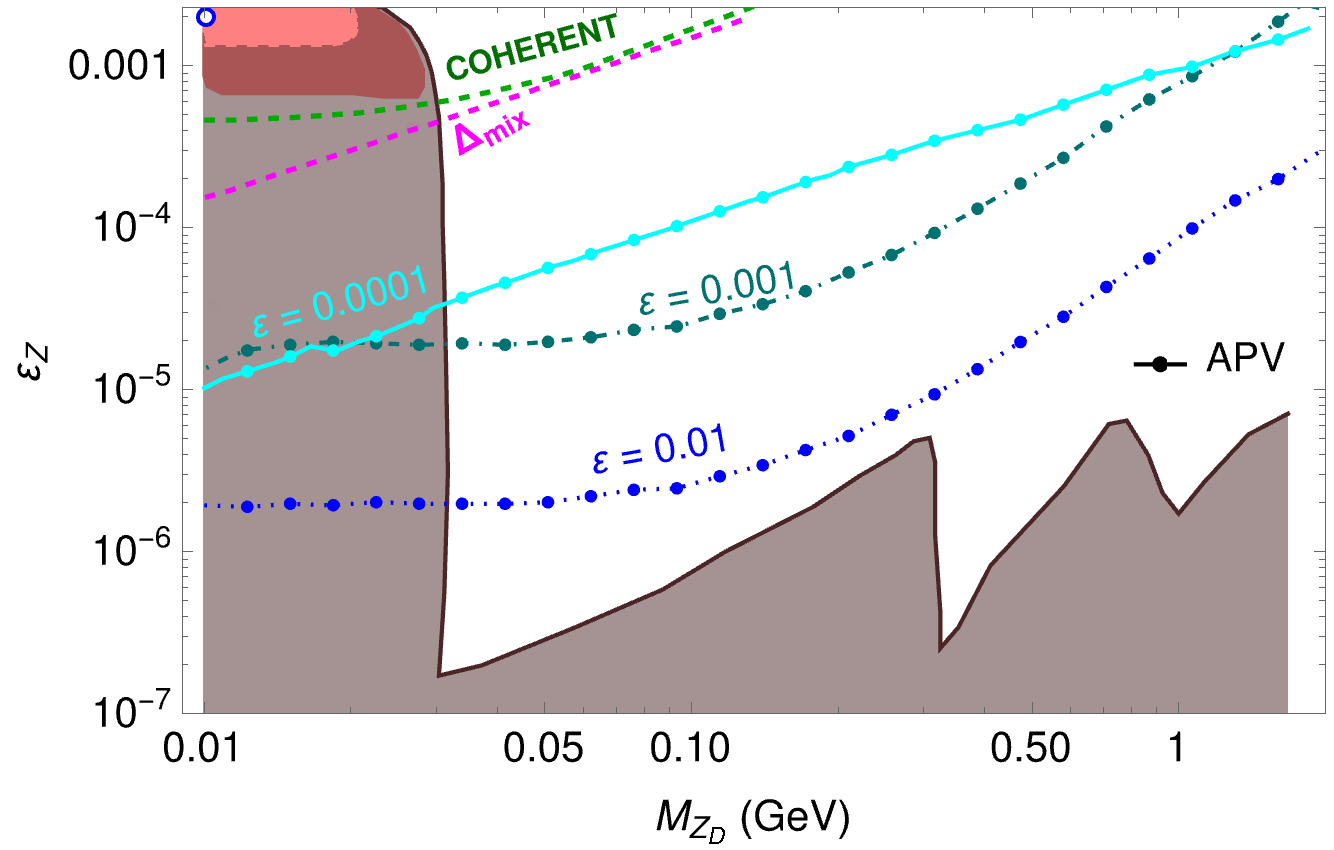}~~
    \includegraphics[scale=0.4]{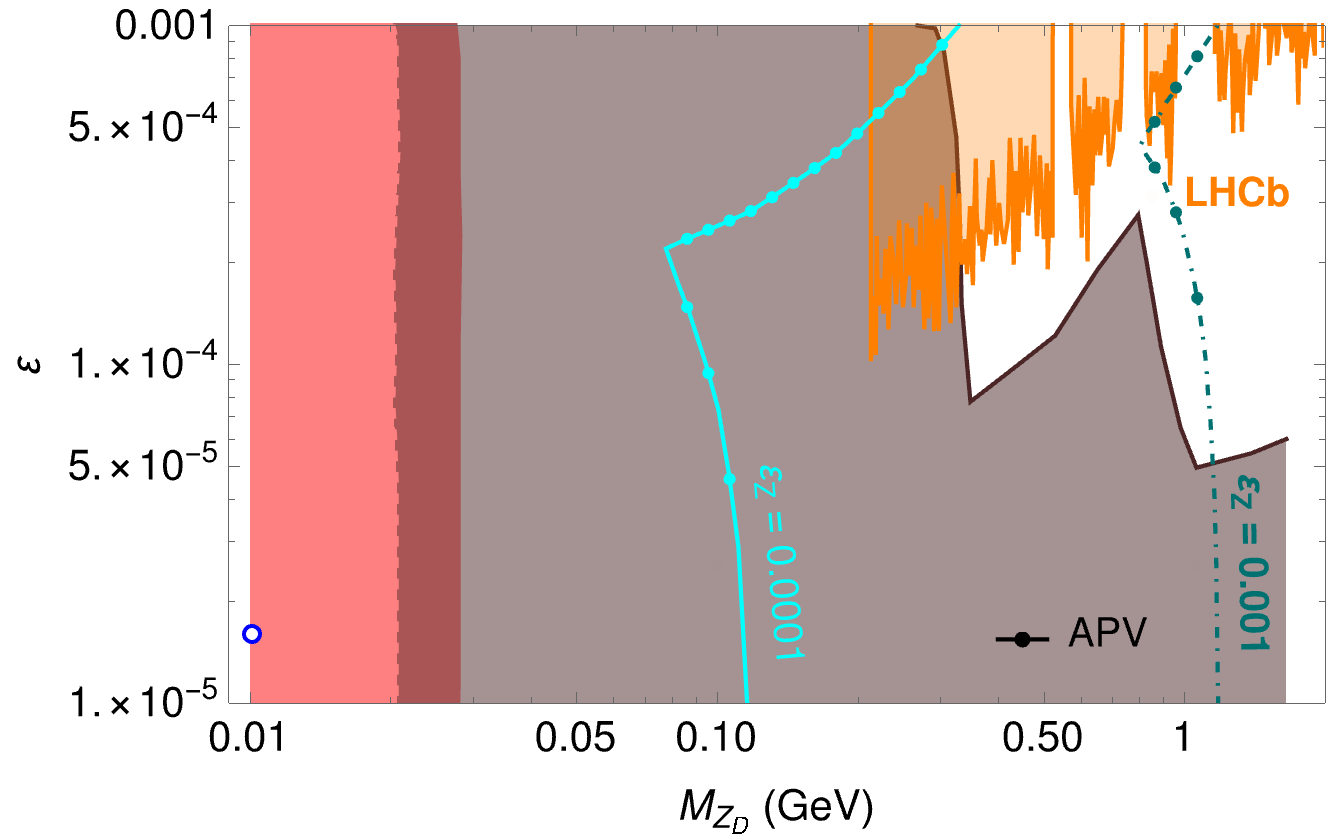}
    \caption{The $1\sigma$ (pink), $2\sigma$ (brown) and 3$\sigma$ (dark brown) regions allowed by the data in Table~\ref{tab:data} for Case A. The best fit point is marked by the blue circle. Top panel: $2\sigma$ upper limit from $B_s^0$-$\overline{B}_s^0$ mixing and $1\sigma$ upper limit from COHERENT neutrino scattering data are shown by the dashed magenta and green curves, respectively. The $3\sigma$ upper limits on $\eps_Z$ from the APV measurements for $\eps = 0.0001, 0.001$ and $0.01$ are shown by the cyan, dark cyan and blue dotted curves, respectively.
    Bottom panel: The orange shaded region is excluded by LHCb dark photon searches at the 90\% CL. The $3\sigma$ upper limits on $\eps$ from APV measurements for $\eps_Z = 0.0001$ and $0.001$ are shown by the cyan and dark cyan curves, respectively; regions to the left of the curves are excluded.}
    \label{fig:CaseA}
\end{figure}

\begin{figure}[h]
\centering
    \includegraphics[scale=0.7]{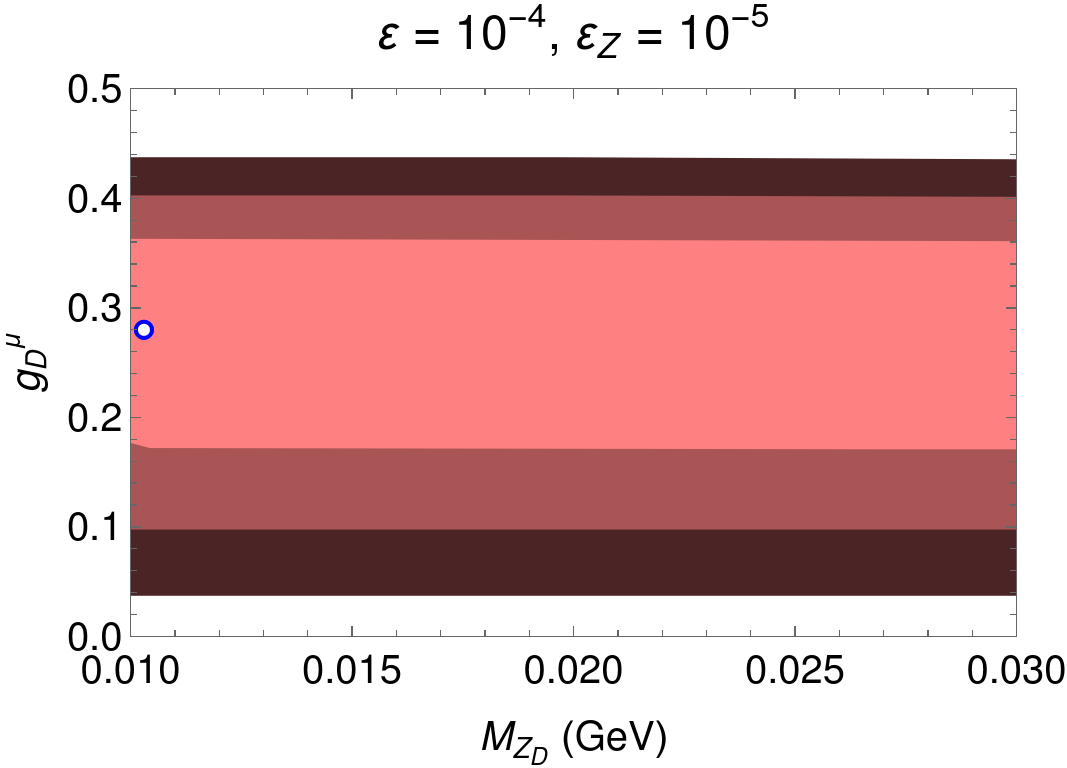}
    \caption{The $1\sigma$, $2\sigma$ and $3\sigma$ allowed regions for Case B with the best fit point marked by a blue circle. However, the entire parameter space is ruled out by
    measurements of $K^+ \to \mu^+ \nu X$,  $X = invisible/e^+ e^-$)  and separately by the $W$ boson width.}
    \label{fig:CaseB}
\end{figure}

\begin{figure}[h]
    \centering
    \includegraphics[scale=0.8]{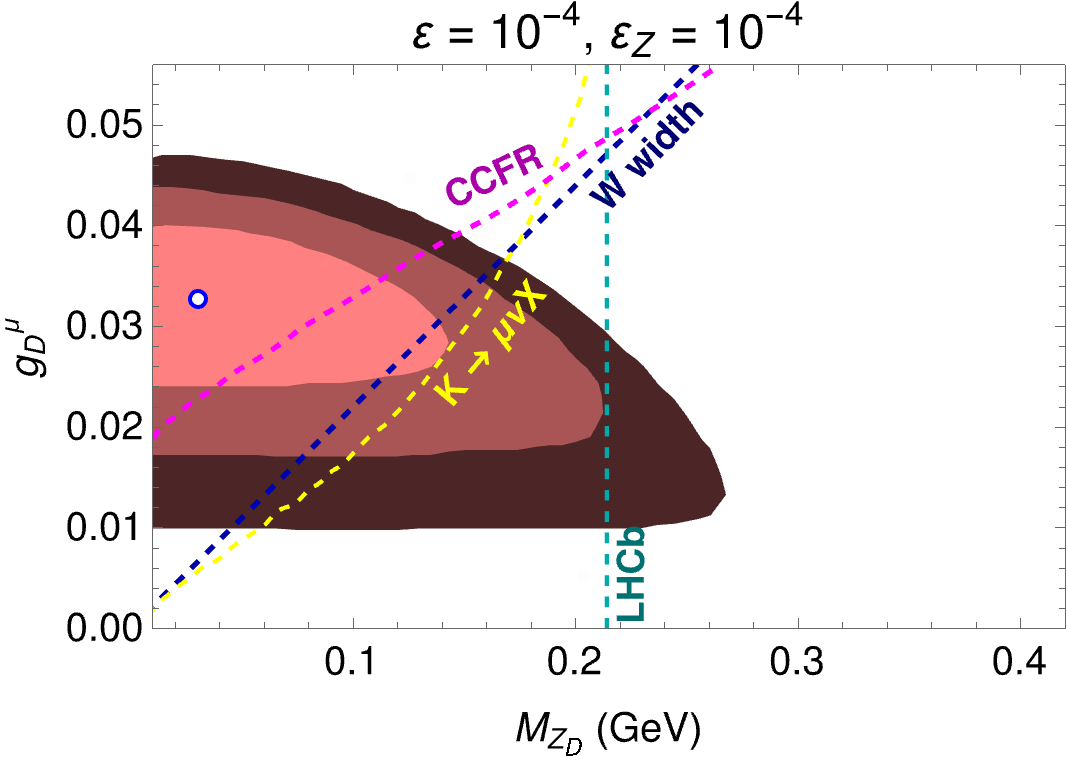}
    \caption{The $1\sigma$, $2\sigma$ and $3\sigma$ allowed regions for Case C with the best fit point marked by a blue circle. 
    Upper limits from neutrino trident production at CCFR (at 95\% CL), $K \to \mu \nu X$ (at 90\% CL) and the $W$ width (at 95\% CL) are shown by the dashed magenta, yellow and dark blue curves, respectively. 
    Dark photon searches at LHCb rule out the region to the right of the vertical dashed dark cyan line at 90\% CL.}
    \label{fig:CaseC}
\end{figure}

%%%%%%%%%%%%%%%%%%%%%%%%%%%%%%%%%%%%%
\section{Muon anomalous magnetic moment}
%%%%%%%%%%%%%%%%%%%%%%%%%%%%%%%%%%%%%%%

The anomalous magnetic moment of the muon, \( a_\mu = (g-2)_\mu / 2 \), shows a discrepancy between experimental measurements and the Standard Model (SM) expectation. The difference, denoted \( \Delta a_\mu \), is approximately $251(59) \times 10^{-11}$. In Case A, the dark boson \( Z_D \) couples to muons via mixing, while in Cases B and C, \( Z_D \) couples directly to muons. The new contributions to \( a_\mu \) in these scenarios vary.

In Case A, the contribution is significantly affected by the axial-vector interaction, leading to a large enhancement in \( a_\mu \). In Case B and Case C, the contribution is less pronounced but still notable. The values for \( a_\mu \) at the best-fit points are \( a_\mu^A = -3.45 \times 10^{-7} \), \( a_\mu^B = 0.00077 \), and \( a_\mu^C = 7.38 \times 10^{-6} \), all of which exceed the measured discrepancy from the SM. This suggests the need for additional new physics, such as a dark charged scalar, which could provide a negative contribution to \( a_\mu \) and help reconcile the results.

\section{Summary}
This study investigates the contributions of a dark photon and dark $Z$ boson to $b \to s \ell^+ \ell^-$ decays, focusing on their effects on the decay amplitudes and incorporating hadronic decays of the dark boson. By fitting to experimental data, we estimate the mass and mixing parameters of the dark boson.

The base model, where the dark $Z$ boson does not directly couple to charged leptons, is ruled out by low-energy experimental constraints. Two extensions of the model, where the dark $Z$ couples directly to muons or electrons, improve the fit to the data. However, the viable parameter space is significantly constrained by other experimental results, especially the anomalous magnetic moment of the muon.

In Case A, the model requires a dark boson mass of less than 30 MeV and specific mixing parameters, but the parameter space is excluded by atomic parity violation experiments. For larger masses, stringent constraints from flavor-changing neutral currents limit the model further.

Case B adds a direct muon coupling, but this results in the exclusion of the entire parameter space due to enhancements in processes like $K \to \mu \nu X$ and the $W$ boson width.

Case C refines the model by introducing a fine-tuned electron coupling, which cancels the electron-mixing contribution, allowing the model to bypass previous constraints. A small viable region remains, but reconciling this with the muon anomalous magnetic moment requires additional new physics.

\bibliographystyle{JHEP}
\bibliography{ref}

\end{document}